\documentclass[12pt]{article}

\usepackage{a4wide}

\usepackage{amssymb}
\usepackage{latexsym}

\usepackage{epsfig}
\usepackage{graphicx}

\usepackage{amsmath,amsthm,amsfonts,amssymb}

\def\de{\partial}
\def\a{\alpha}
\def\b{\beta}

\def\d{\delta}

\def\l{\lambda}
\def\k{\kappa}
\def\m{\mu}
\def\n{\nu}

\def\th{\theta}

\newcommand{\be}{\begin{equation}}
\newcommand{\ee}{\end{equation}}
\newcommand{\bea}{\begin{eqnarray}}
\newcommand{\eea}{\end{eqnarray}}
\newcommand{\beqar}{\begin{eqnarray*}}
\newcommand{\eeqar}{\end{eqnarray*}}
\newcommand{\eg}{{\it e.g.,}\ }
\newcommand{\ie}{{\it i.e.,}\ }

\newcommand{\nn}{\nonumber}

\begin{document}

\begin{titlepage}

\rightline{hep-th/0501112}
\rightline{January 2005}

\begin{centering}
\vspace{2cm}
{\large {\bf Brane-bulk matter relation for a purely conical\\ codimension-2 brane world }}\\

\vspace{1.5cm}

 {\bf Eleftherios~Papantonopoulos}$^{a,*}$ and {\bf Antonios~Papazoglou} $^{b,**}$ \\
\vspace{.2in}

$^{a}$ National Technical University of Athens,\\ Physics
Department,\\ Zografou Campus, GR 157 80, Athens, Greece. \\
\vspace{3mm}
$^{b}$ \'Ecole Polytechnique F\'ed\'erale de Lausanne,\\ Institute of Theoretical Physics,\\
SB ITP LPPC BSP 720, CH 1015, Lausanne, Switzerland.

\end{centering}
\vspace{3cm}

\begin{abstract}

We study gravity on an infinitely thin codimension-2 brane world,
with purely conical singularities and in the presence of an
induced gravity term on the brane. We show that in this
approximation,  the energy momentum tensor of the bulk is strongly
related to the energy momentum tensor of the brane and thus the
gravity dynamics on the brane are induced by the bulk content.
This is in contrast with the gravity dynamics on a codimension-1
brane. We show how this strong result is relaxed  after
including a Gauss-Bonnet term in the bulk.

\end{abstract}

\vspace{3cm}
\begin{flushleft}

$^{*}~$ e-mail address: lpapa@central.ntua.gr \\
$ ^{**}$ e-mail address: antonios.papazoglou@epfl.ch

\end{flushleft}
\end{titlepage}

 \section{Introduction}

Recently, there have been many observational and theoretical
motivations for the study of theories with extra spacetime dimensions
and in particular the braneworld scenario. From the observational
side, the current paradigm, supported
 by many recent observations \cite{Spergel:2003cb}, is that most of the energy
content of our universe is of the form of dark matter and dark
energy. Although there have been many plausible explanations for
these dark components, it is challenging
 to try to explain these exotic ingredients of the universe using alternative gravity
 theories as such of the braneworlds. From the theoretical side, such extra-dimensional
 braneworld models are ubiquitous in theories like string or M-theory. Since
 these theories claim to give us a fundamental description of nature, it is important to study
what kind of gravity dynamics they predict.  The hope is to
propose such modified gravity theories, which share many common
features with general relativity, but
 at the same time give alternative non-conventional cosmology.

The essence of the braneworld scenario is that the Standard
Model, with its matter and gauge interactions, is localized on a
three-dimensional hypersurface (called brane) in a
higher-dimensional spacetime. Gravity propagates in all spacetime
 (called bulk) and thus connects the Standard Model sector with the internal space
dynamics. This idea, although quite old \cite{earlybranes},
gained momentum
 the last years \cite{AADD,randall} because of its connection with string theory.  The
 cosmology of this and other related models
with one transverse to the brane extra dimension (codimension-1
brane models) is well
 understood (for a review see \cite{reviews}). In the cosmological generalization of \cite{randall},
 the early times (high energy limit) cosmological evolution is modified
by the square of the matter density on the brane, while the bulk
leaves its imprints on the brane by the ``dark radiation" term
 \cite{RScosmology}. The presence of a bulk cosmological constant
 in \cite{randall} gives conventional
cosmology at late times (low energy limit) \cite{RScosmology}.

In the above models there are strong theoretical arguments for
including in the gravitational action extra curvature terms in addition to  the higher dimensional
 Einstein-Hilbert term. The localized matter fields on the brane, which couple
to bulk gravitons, can generate via quantum loops  a localized
four-dimensional
 kinetic term for gravitons~\cite{oldinduced}. The latter comes in the gravitational
action as a four-dimensional scalar curvature term localized at
the position of the brane
   (induced gravity)~\cite{induced} \footnote{There has been
   a lot of discussion about the potential problem of the extra polarization states
   of the massive  gravitons regarding phenomenology (discontinuity problem),
   but the particular model seems to be consistent in a non-trivial way \cite{indconsistency}.}.
   In addition, curvature square terms in the bulk, in the  Gauss-Bonnet
 combination,  give the most general action with second-order field equations in five
dimensions~\cite{Lovelock:1971yv}. This correction is also
motivated by string theory, where
  the Gauss-Bonnet term corresponds to
the leading order quantum correction to gravity, and its presence
guarantees a ghost-free action~\cite{GB}. Let us note, however,
that if the curvature squared terms are to play an important role
in the low energy dynamics, one may face a difficulty when
interpreting gravity as an effective field theory (in the sense
that even higher dimensional operators would seem to be also
relevant).

If curvature corrections are included in \cite{randall}, the
presence of the 3-brane gives similar modifications to the
standard cosmology. In the case of a pure Gauss-Bonnet term in the
bulk, the early times cosmology is modified by a term proportional
to the matter density on the brane to the power two thirds
 \cite{GBcosmology}. If an induced gravity term
is included, the conventional cosmology is
modified by the square root of the matter density on the brane at
low energies \cite{indcosmology}. Finally, if both curvature corrections are present,
the cosmological evolution is affected mainly by the induced
gravity corrections at late times.

In six dimensions and for codimension-2  braneworlds, the gravity
dynamics appear even more radical and still a good understanding
of cosmology and more generally gravity in such theories is
missing. The most attractive feature of codimension-2 braneworlds
is that the vacuum  energy (tension) of the brane
 instead of curving the brane world-volume, merely induces a deficit angle in the
 bulk solution around the brane \cite{Chen:2000at} (see also \cite{stingdefects} for string-like defects in six dimensions). This looks very promising in relation to the cosmological
constant problem, although in existing models, nearby curved
solutions cannot be
 excluded \cite{6d}, unless one allows for singularities more severe than conical
  in particular supersymmetric models \cite{6dsusy}.  It was soon realized \cite{Cline:2003ak}
   that one can only find nonsingular
solutions if the brane energy momentum tensor is proportional to
its induced metric, which means simply that it is pure tension. A
non-trivial energy momentum tensor on the brane  causes
singularities in the metric around the braneworld, which
necessitates the introduction of a cut-off (brane thickness)
\cite{thickadd,Vinet:2004bk,Navarro:2004di}.

An alternative approach to study the gravitational dynamics of
matter on infinitely thin branes is to modify the gravitational
action as discussed previously. Indeed, it was shown in
\cite{Bostock:2003cv} that  the inclusion of a Gauss-Bonnet term
in the gravitational  action allows a non-trivial energy momentum
tensor on the brane, and in the thin brane limit,  four
dimensional gravity is recovered as the dynamics of the induced
metric on the brane. The peculiar characteristic of this way to
obtain four dimensional gravity for codimension-2 branes, is that,
apart from the inclusion of a (deficit angle independent)
cosmological constant term, there appear to be no corrections to
the Einstein equations coming from the extra dimensions in the
purely conical case. Another possibility, discussed in
\cite{intersections}, is to study (instead of conical 3-branes)
codimension-2 branes sitting at the intersection of codimension-1
branes  in the presence again of a bulk Gauss-Bonnet term.

In this paper, we show that results similar to the Gauss-Bonnet
case, {\it i.e.} four dimensional gravity for an arbitrary energy
momentum tensor,  can be obtained if we include in the action an
induced gravity correction term instead. Again, in the purely
conical case, there appear to be no corrections to the Einstein
equations coming from the extra dimensions. The most important
observation is that the brane and bulk energy momentum tensors are
strongly  related and any cosmological evolution on the brane is
dictated by the bulk content. We also see how this correlation is
relaxed in the case where bulk Gauss-Bonnet terms and brane
induced gravity terms are combined. Thus, the necessary presence
of extra curvature terms in the gravitational action in order to
give non-trivial gravitational dynamics on a codimension-2 brane,
leads to a realistic cosmological evolution on the
 brane in the thin brane limit, only if a Gauss-Bonnet term is 
included. However, let us note that the most physical way to investigate
  the dynamics of codimension-2 branes, is by giving thickness to the brane  \cite{Vinet:2004bk,Navarro:2004di}.

The paper is organized as follows. In Sec.2 we present a
six-dimensional induced gravity model and we discuss its
gravitational dynamics on the brane. In Sec.3 we compare our
results with the five-dimensional induced gravity case and discuss
their differences. In Sec.4 we extent our analysis including the
Gauss-Bonnet term in the bulk and finally in Sec.5 we summarize
our results and conclude.

\section{Induced gravity in six dimensions}

We consider a six-dimensional theory with general bulk dynamics encoded in a Lagrangian ${\cal L}_{Bulk}$ and a
3-brane at some point $r=0$ of the two-dimensional internal space with general dynamics ${\cal L}_{brane}$ in its world-volume. If we include an induced curvature term localized at the position of the brane, the total action is written as:

\bea
 {\cal S}&=&{M^4_6 \over 2}\left[\int d^6x \sqrt{G}R^{(6)}+r_c^2\int d^4x
\sqrt{g}R^{(4)} ~{\d(r) \over 2 \pi L}\right]\nn \\  &+& \int d^6x
{\cal L}_{Bulk} + \int d^4x {\cal L}_{brane}~{\d(r) \over 2 \pi
L}~.\label{inaction} \eea

In the above action, $M_6$ is the six-dimensional Planck mass,
 $M_4$ is the four-dimensional one and $r_c=M_4/M_6^2$ the cross over scale between four-dimensional and six-dimensional gravity.
 The above induced term has been
written in the particular coordinate system in which the metric is

\be ds_6^2=g_{\m\n}(x,r)dx^\m
dx^\n+dr^2+L^2(x,r)d\th^2~,
\ee
where $g_{\mu\nu}(x,0)$ is the
braneworld metric and $x^{\mu}$ denote four non-compact
dimensions, $\mu=0,...,3$, whereas $r,\th$ denote the radial and
angular coordinates of the two extra dimensions (the r direction may or may not  be compact and the $\th$ coordinate ranges form $0$ to $2\pi$). 
Capital $M$,$N$ indices will take values in the six-dimensional
space. Note, that we have assumed that there exists an azimuthal
symmetry in the system, so that both the induced four-dimensional
metric and the function $L$ do not depend on $\th$. The normalization of the $\d$-function is the one discussed in \cite{Leblond:2001xr}.

To obtain the
braneworld equations we expand the metric around the brane as

\be
L(x,r)=\beta(x)r+O(r^{2})~.
\ee

At the boundary of the internal
two-dimensional space where the 3-brane is situated, the function $L$
behaves as $L^{\prime}(x,0)=\beta(x)$, where a prime denotes
derivative with respect to $r$. As we will see in the following, the demand that the space in the vicinity of the conical singularity is regular, imposes the
supplementary conditions that $\de_\m \b=0$ and  $\partial_{r}g_{\mu\nu}(x,0)=0$.

The Einstein equations which are derived from the above action in
the presence of the 3-brane are

\be
 G^{(6)N}_M + r_c^2
G^{(4)\n}_\m \d_M^\m \d^N_\n {\d(r) \over 2 \pi L}={1 \over M_6^4}
\left[T^{(B)N}_M+T^{(br)\n}_\m \d_M^\m \d^N_\n {\d(r) \over 2 \pi
L}\right]~,
\label{einsequat}
\ee
with $G^{(6)N}_M$ and $G^{(4)\n}_\m$ the six-dimensional and the four-dimensional Einstein tensors respectively,   $T^{(B)N}_M$ the bulk energy
momentum tensor and $T^{(br)\n}_\m$ the brane one.

The six-dimensional Ricci tensor components can be written in
terms of the four-dimensional ones, the extrinsic curvature $K_{\m\n}$ and
the $L$ function as \cite{Navarro:2004di}

 \bea
 R_\m^{(6)\n}&=&-{(\sqrt{g}LK_\m^\n)'
\over 2\sqrt{g}L}+R^{(4)\n}_\m-
{\nabla^{(4)}_\m \de^\n L \over L}~,\label{rmn}\\
R_\th^{(6)\th}&=&-{(\sqrt{g}L')' \over \sqrt{g}L}-{\square^{(4)}L \over L}~,\label{rthth}\\
R_r^{(6)r}&=&-{L'' \over L}-{1 \over 2}K'-{1 \over 4}K_\m^\n K^\m_\n~,\label{rrr}\\
R^{(6)}_{\m r}&=&-{\de_\m L' \over L}+{K_\m^\n\de_\n L  \over 2 L
}+{1 \over 2}\nabla_{(4)}^\n(K_{\m\n}-g_{\m\n}K)\label{rmr}~.
\eea

 The
extrinsic curvature in the particular gauge $g_{rr}=1$ that we are considering,  is given by $K_{\m\n}=g'_{\m\n}$.
 The above decomposition will be helpful in the following for finding the induced dynamics on the brane.

We will now use the fact that the second derivatives of the metric
functions contain $\d$-function singularities at the position of
the brane. The nature of the singularity then gives the following
relations \cite{Bostock:2003cv}

\bea
{L'' \over L}&=&-(1-L'){\d(r) \over L}+ {\rm non-singular~terms}~,\\
{K'_{\m\n} \over L}&=&K_{\m\n}{\d(r) \over L}+ {\rm
non-singular~terms}~.
\eea

From the above singularity expressions and the decomposition (\ref{rmn})-(\ref{rrr}),
we can  match the singular parts of
the Einstein equations (\ref{einsequat}) and get the following
``boundary" Einstein equations

\be
G^{(4)\n}_\m|_0 = {1 \over r_c^2
M_6^4}T^{(br)\n}_\m +{2\pi \over r_c^2} (1-\b)\d_\m^\n +{2 \pi L
\over 2r_c^2}(K_{\m}^\n-\d_\m^\n K)|_0 \label{indeq}~.
\ee
where we denote by $|_0$ the value of the corresponding function at $r=0$.

We will now make the assumption that the singularity is purely
conical. In the opposite case, there would be curvature
singularities $R^{(6)} \propto 1/r$, because in the Ricci tensor
$R^{(6)}_{\mu\nu}$ there are terms of the form \cite{Bostock:2003cv}

\be
R^{(6)}_{\mu\nu}=-\frac{1}{2}\frac{L^{\prime}}{L}\partial_{r}g_{\mu\nu}+...=-\frac{\partial_{r}g_{\mu\nu}}{2r}
+\mathcal{O}(1)~,
\ee
 which in the vicinity of $r=0$ are singular
if $\partial_{r}g_{\mu\nu}(x,0)\neq0$. The absence of this type of
singularities imposes the requirement that
$K_{\m\n}|_0=0$. Then, the Einstein equations
(\ref{indeq}) reduce to

\be
G^{(4)}_{\m\n}|_0= {1 \over r_c^2 M_6^4}T^{(br)}_{\m\n} +{2 \pi
\over r_c^2} (1-\b)g_{\m\n}|_0 \label{conindeq}~.
\ee

The four-dimensional Einstein equations (\ref{conindeq}) describe
the gravitational dynamics on the brane. The effective four-dimensional Planck mass and cosmological constant are simply

\bea
 M^{2}_{Pl}~=~M^2_4&=& r_c^2 M_6^4~,\\
\Lambda_{4}&=&\lambda-2\pi M_6^4 (1-\b)
\eea
where $\lambda$ is the contribution of the vacuum energy of the brane fields. The normalization of $\Lambda_4$ is defined by the convention that the four dimensional Einstein equation reads \mbox{$G_{\m\n}={1 \over M_{Pl}^2}(T_{\m\n}-\Lambda_4 g_{\m\n})$}. Note that in contrast to the case of \cite{Bostock:2003cv}, the four dimensional Planck mass is independent of the deficit angle.

 Furthermore, it is interesting to see  that contrary to
the five-dimensional case, the induced gravity term in six
dimensions does not introduce any correction terms, apart from a cosmological term,  in the
four-dimensional Einstein equations on the brane, unless singularities of
other type than conical  are allowed in the theory, and a
regularization scheme is employed. In the latter case, the last term of the right hand side of (\ref{indeq}) would provide information of the bulk physics. 
This absence of corrections in the purely conical case,  is exactly what happens also in the case of the bulk Gauss-Bonnet theory \cite{Bostock:2003cv}.

What is important to note at this point, is that although we have found a ``boundary'' Einstein equation,
there is more information about the dynamics of the theory  contained in the full six-dimensional Einstein equations.  As we will shortly see,  the consistency of the Einstein equations (\ref{einsequat}) will give us information about the matter content of the brane.

Firstly,  the $(\m r)$ component of the Einstein  equations (\ref{einsequat}) evaluated at $r=0$ and  with the help of  ({\ref{rmr})  gives

\be
 \left.{\de_\m L' \over L}\right|_0=-{1 \over M_6^4}
T_{\m r}|_0~. \label{tmr}
\ee

This equation is consistent only
for $\b=L'|_0={\rm const.}$, because otherwise $\de_\m L' / L$ will have  an
$1/r$ singularity. This also means that ${\square^{(4)}L
\over L}|_0=0$. This does not exclude, however, the possibility that there is $x$ dependence of $L$ close to the brane. Only the linear term in $r$ of the $L$ expansion is constant.

Evaluating the $(rr)$ component of the Einstein equation (\ref{einsequat}) at the position of the brane $r=0$ with the help of (\ref{rmn})-(\ref{rrr}), we get

\be
R^{(4)}|_0=-{2 \over M_6^4}T^{(B)r}_r|_0~.\label{r4}
\ee

On the other hand, the trace of the  ``boundary'' Einstein equation (\ref{conindeq}) gives us the relation

\be
 R^{(4)}|_0=-{1 \over r_c^2 M_6^4}T^{(br)\m}_\m-{8 \pi \over
r_c^2}(1-\b)~,\label{trace}
\ee

Combining now equations (\ref{r4}) and (\ref{trace}) we arrive at a relation between
 the bulk and
brane energy momentum tensors

 \be
T^{(B)r}_r|_0={1 \over
2r_c^2}\left[T^{(br)\m}_\m+8 \pi M_6^4
(1-\b)\right]~.\label{bulkbranem}
\ee

This equation is the central result of our paper. It  constitutes a very
strong tuning relation between brane ($T^{(br)\m}_\m$) and bulk ($T^{(B)r}_r|_0$) matter. It shows that, in
order to have some cosmological evolution on the brane (\ie time
dependent $T^{(br)}_{\m\n}$) and since $\b$ is constant,  the bulk content should evolve as well in a precisely tuned way.

 In
other words, one can say that the brane cosmological evolution is
induced by the bulk content. The ``boundary'' Einstein equation may not contain any bulk information at first sight, but the  admissible brane energy momentum tensor is dictated by the bulk energy momentum tensor.   It is difficult to justify why such a relation would be physically natural.

\section{Comparison with five dimensions}

In this section we will compare the above result with what is
happening in five dimensions. The action of a general
five-dimensional theory with a 3-brane at the point $r=0$ of the
extra dimension, and with an induced curvature term localized on
it, is

\bea
 {\cal S}&=&{M^3_5 \over 2}\left[\int d^5x \sqrt{G}R^{(5)}+r_c\int d^4x
\sqrt{g}R^{(4)} ~\d(r)\right]\nn \\  &+& \int d^5x {\cal L}_{Bulk}
+ \int d^4x {\cal L}_{brane}~\d(r) \label{action}~.
\eea

 In the
above action, $M_5$ is the five-dimensional Planck mass, $M_4$ is
the four dimensional one and $r_c=M_4^2/M_5^3$ the cross over
scale of the five-dimensional theory. The above induced term has
been written in the particular coordinate system in which the
metric is written as \be ds_5^2=g_{\m\n}(x,r)dx^\m dx^\n+dr^2~,
\ee where the $x^{\mu}$ denote the usual four non-compact
dimensions, $\mu=0,...,3$, whereas $r$ denotes the radial extra
coordinate and capital $M$,$N$ indices will now take values in the
five-dimensional space.

The Einstein equations which are derived from the action
(\ref{action}) in the presence of the 3-brane are

\be
 G^{(5)N}_M +
r_c G^{(4)\n}_\m \d_M^\m \d^N_\n \d(r)={1 \over M_5^3}
\left[T^{(B)N}_M+T^{(br)\n}_\m \d_M^\m \d^N_\n
\d(r)\right]~,\label{5deineq}
\ee
 with $T^{(B)N}_M$ the bulk
energy momentum tensor and $T^{(br)\n}_\m$ the brane one. The
five-dimensional Ricci tensor components can be written in terms
of the four-dimensional ones and the extrinsic curvature as

\bea
R_\m^{(5)\n}&=&-{(\sqrt{g}K_\m^\n)' \over 2\sqrt{g}}+R^{(4)\n}_\m~,\label{5rmn}\\
R_r^{(5)r}&=&-{1 \over 2}K'-{1 \over 4}K_\m^\n K^\m_\n~,\label{5rrr}\\
R^{(5)}_{\m r}&=&{1 \over
2}\nabla_{(4)}^\n(K_{\m\n}-g_{\m\n}K)~.\label{5rmr}
\eea

 The extrinsic
curvature is $K_{\m\n}=g'_{\m\n}$ in the particular gauge we are
using. We assume that there is a $Z_2$ symmetry around $r=0$, so
that we can write the extrinsic curvature as

\be
K'_{\m\n} =
K_{\m\n} \d(r) + {\rm non-singular~terms}~.
\ee

Then, matching the singular parts of the Einstein equations
(\ref{5deineq}) using (\ref{5rmn})-(\ref{5rmr}) we get the
following ``boundary" Einstein equations

\be
G^{(4)\n}_\m|_0 = {1
\over r_c M_5^3}T^{(br)\n}_\m +{1 \over 2r_c}(K_{\m}^\n-\d_\m^\n
K)|_0~.
\ee

These four-dimensional Einstein equations describe the
gravitational dynamics on a codimension-1 brane with induced
gravity, and the extrinsic curvature terms in the right hand side of the
equations can be considered as corrections to the four-dimensional
standard Einstein equations.

We can get further information evaluating the Einstein equations
(\ref{5deineq}) at the position of the brane $r=0$. The $(\m r)$ component of the Einstein  equation (\ref{5deineq}) evaluated at $r=0$, with the help of
(\ref{5rmr})  gives

\be
\nabla_{(4)}^\n(K_{\m\n}-g_{\m\n}K)|_0={2 \over M_5^3}T^{(B)}_{\m r}~.
\ee

There is no condition coming from this equation, in contrast with
what is happening in six dimensions, in which the $T^{(B)}_{\m r}$
component of the bulk energy momentum tensor is restricted. The $(r r)$ component of the Einstein  equation (\ref{5deineq}) evaluated at $r=0$, with the help of (\ref{5rrr}) gives

\be
 R^{(4)}|_0+{1
\over 4}(K_\m^\n K^\m_\n-K^2)|_0=- {2 \over M_5^3}T^{(B)r}_r|_0~.
\ee

From the above equation we see that the bulk matter content does
not necessarily dictate the brane cosmological evolution. This is
because the extrinsic curvature on the brane $K_{\m \n}$ can be
non-trivial and it is this one which plays the most crucial role
in the cosmology. In other words, in five dimensions it is the
freedom of the brane to bend in the extra dimension which makes
the evolution not tuned to the bulk matter content. The absence of
such bending in six dimensions (imposed by singularity arguments)
gives the bulk the crucial role for how the brane evolves.

\section{Inclusion of a Gauss-Bonnet term}

In this section we will introduce a Gauss-Bonnet term in the
six-dimensional action (\ref{inaction}) and see how the above
results are modified. In this case the action (\ref{inaction}) is
augmented by the term

\be S_{GB}={M_6^4 \a \over 2}\int d^6x
(R^{(6)~2}-4R^{(6)~2}_{MN}+R^{(6)~2}_{MNK\Lambda})~.
\ee

 Then the
variation of the above action introduces an extra term in the left hand side of the
Einstein equations (\ref{einsequat}),

\bea
 H_M^N&=&-\a \left[{1
\over 2}\d_M^N (R^{(6)~2}
-4R^{(6)~2}_{K\Lambda}+R^{(6)~2}_{ABK \Lambda})\right.-2R^{(6)}R^{(6)N}_{M}\nonumber\\
&&+4R^{(6)}_{MP}R^{NP}_{(6)}\phantom{{1 \over 2}}~\left.
+4R^{(6)~~~N}_{KMP}R_{(6)}^{KP} -2R^{(6)}_{MK\Lambda
P}R_{(6)}^{NK\Lambda P}\right]~.
\eea

Equating the singular terms of the Einstein equations by the
standard procedure of section 2, and demanding that the singularity is purely
conical, we obtain the following
 ``boundary" Einstein equations

 \be
G^{(4)}_{\m\n}|_0={1 \over M_6^4 (r_c^2+8\pi
(1-\b)\a)}T^{(br)}_{\m\n}+{2\pi (1-\b) \over r_c^2+8\pi
(1-\b)\a}g_{\m\n}|_0 \label{einsteincomb}~.
\ee

 Equation
(\ref{einsteincomb}) describes the gravitational dynamics on a
codimension-2 brane when both induced gravity and Gauss-Bonnet
correction terms are present. The effective four-dimensional Planck mass and cosmological constant are simply

\bea
 M^{2}_{Pl}&=&M_6^4 (r_c^2+8\pi (1-\b)\a)~,\\
\Lambda_{4}&=&\lambda-2\pi M_6^4 (1-\b)~,
\eea
where $\lambda$ is the brane tension.  Note that the Planck mass this time can depend on the deficit angle. This is an effect of solely the bulk Gauss-Bonnet term.

Evaluating the the $(rr)$ component of the Einstein equation at the position of the brane $r=0$ we obtain the following relation

 \be
 R^{(4)}|_0+\a
(R^{(4)~2}-4R^{(4)~2}_{\k\l}+R^{(4)~2}_{\a\b\k\l})|_0 =-{2 \over
M_6^2}T^{(B)r}_r|_0
\label{rrcomb}~.
\ee

From (\ref{rrcomb}) we see that there can be no relation between the extra dimensional component $T^{(B)r}_r|_0$ of the bulk energy momentum tensor  at the position of the brane with the brane energy momentum tensor $T^{(br)\m}_\m$. This is due to the appearance of the Riemann curvature, which cannot be evaluated from previous equations (the Ricci tensor and scalar can be substituted from (\ref{einsteincomb})). Instead, using  (\ref{einsteincomb}) and
(\ref{rrcomb}), one can {\it solve} for $R^{(4)~2}_{\a\b\k\l}|_0$  as a function of the brane and bulk matter at the position of the brane.

\section{Conclusions}

In this work, we have considered the gravitational dynamics of conical codimension-2 branes
 of infinitesimal thickness. The assumption of the absence of metric singularities
  more severe than conical, imposes a strong constraint in the allowed matter on the brane.
   In particular we have considered theories with codimension-2 branes which are augmented
    with an induced gravity term on the brane and a Gauss-Bonnet term in the bulk.
     These additions where known in the literature to give enough
      freedom for the brane dynamics to admit general matter.

The crucial observation of this paper is that the dynamics of the latter theories  is not exhausted
 by studying the ``boundary'' Einstein equation, which is exactly four-dimensional
  and bears no information of the internal space (modulo a cosmological constant contribution).
   In the case of a pure brane induced gravity term, the higher dimensional Einstein equations evaluated at the position of the brane,
    give a very precise and strong relation between the matter on the brane
     and the matter in the bulk in the vicinity of the brane. So, for example,
      in a cosmological setting, in order to have a cosmological
       evolution of general energy density and pressure on the brane,
        the bulk matter should organize itself so that the above-mentioned brane-bulk matter relation is satisfied.
         In other words, the bulk energy content is the primary factor for the cosmological evolution on the brane.
           Alternatively, for a static matter distribution on the brane to be possible,
            there should exist its bulk matter ``image''.

This matter constraint is due to the fact that the absence of severe singularities criterion prevents the brane
 to bend in the internal space. The presence of this bending was the reason why in five dimensions the energy
  density on the brane was unrelated to that of the bulk. The latter influenced the cosmological evolution on
   the brane (\eg through the ``dark radiation'' term)  but did not completely determine the evolution
    on the brane as in the six-dimensional case.

This strong relation, that we have noted in this paper, can be avoided with the inclusion of a bulk Gauss-Bonnet term. An even more natural way to achieve this is to relax the requirement of purely conical branes and admit
  general brane solutions with an appropriate regularization (thickening of the brane), 
   so that the singularities are smoothened. In view of the difficulties related to the Gauss-Bonnet term in the context of effective field theory,  this work points out that the thickening of the brane is the most physical
    direction that one should follow in order to discuss the dynamics of codimension-2 branes.

\section*{Acknowledgments}

We are grateful to Hyun Min Lee for pointing out an erroneous consideration  in the first version of the manuscript. A.P. wishes to thank NTUA for hospitality during the course of this work.  This work was supported by the Greek
Education Ministry research program "Pythagoras".

\end{document}